\documentclass{kluwer}    

\newdisplay{guess}{Conjecture}

\begin{document}                                                                                   
\begin{article}
\begin{opening}         
\title{Time-resolved spectroscopy and spectrophotometry of the pulsating
  sdB star PG\,1605+072} 
\author{S.~J. \surname{O'Toole}\email{otoole@physics.usyd.edu.au}}
\author{T.~R. \surname{Bedding}}
\institute{School of Physics, University of Sydney 2006, Australia}
\author{H. \surname{Kjeldsen}}
\institute{Theoretical Astrophysics Center, Aarhus University,
  DK-8000, Aarhus C, Denmark}
\author{M.~A.~S.~G. \surname{J{\o}rgensen}}
\institute{University of Copenhagen, Astronomical Observatory, Juliane
  Maries Vej 30, DK-2100, Copenhagen \O, Denmark}
\runningauthor{O'Toole et al.}
\runningtitle{PG\,1605+072: Latest Results}

\begin{abstract}
We present the latest results from our analysis of time resolved
spectroscopy of PG\,1605+072, a high amplitude pulsating sdB star. The
star is believed to have evolved off the extreme horizontal branch,
and it is uncertain whether it pulsates in $p$- or $g$-modes. Previous
studies of PG\,1605+072 have found amplitude variation and/or very
closely spaced frequencies. We compare our velocity measurements with
new simultaneous spectrophotometric measurements and analyse phase
differences for the 8 frequencies found in velocity. Phase differences
between velocity, spectrophotometry and equivalent width are
discussed.
\end{abstract}

\end{opening}           

\section{PG\,1605+072: a high amplitude sdB pulsator}  

PG\,1605+072 is a high amplitude multiperiodic pulsating sdB
star. Multisite photometry over 15 nights by \inlinecite{ECpaperX}
found up to 55 frequencies between 1.7--5\,mHz, although some may be
artifacts of amplitude variation. Amplitudes in white light are
typically between 1 and 30\,mmag. From spectral analysis,
\inlinecite{HRW99} determined $T_\mathrm{eff}$=32\,300\,K,
log\,$g$=5.25, log\,(He/H)=$-$2.53 and $v\sin i$=39\,km\,s$^{-1}$.

\inlinecite{OBK00} reported the detection of velocity variations in 5
Balmer lines of PG\,1605+072 using small telescopes over 10 nights. In
a more detailed study, \inlinecite{OBK02} detected 8 frequencies in
velocity, with observations taken over 3 months. Some of these may be
artifacts of amplitude variation. \inlinecite{WJP02} presented 32\,h
of 4\,m spectroscopy and detected the dominant modes. Equivalent width
variations in the Balmer lines have also been detected by
\inlinecite{OJK02}, with large amplitude variations between different
Balmer lines.
Here we present rapid spectrophotometry of PG\,1605+072 based on the
spectra of \inlinecite{OBK02}.

\section{Spectrophotometry}

Using the spectra obtained by O'Toole et al. (2002a) we have
calculated three spectrophotometric time-series: one of the entire
spectrum, which we will call ``blue light''; one of the bluest
continuum, which we denote $U^\prime$; and one between H$\gamma$ and
H$\beta$, avoiding the He \textsc{i} 4471\,\AA\ line, which we denote
$B^\prime$. The $U^\prime$ filter is centred on $\sim$3700\,\AA, with
a width of 110\,\AA, while the $B^\prime$ filter is centred on
$\sim$4570\,\AA, with a width of 175\,\AA.

\begin{figure}[htbp]
\vspace{9cm}
\center{\includegraphics{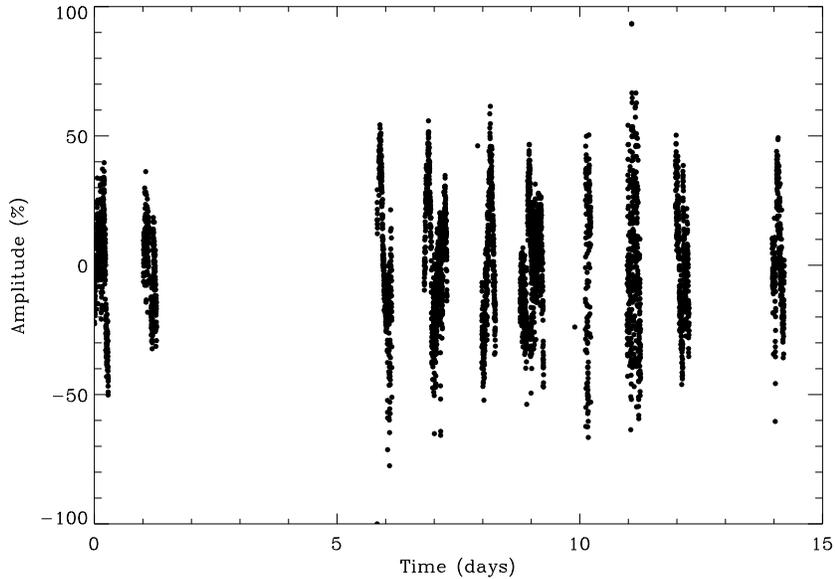}}
\caption{The ``blue'' light curve of PG\,1605+072. Long term
  variations are due variable illumination on the slit. }
\label{fig:bluetime}
\end{figure}

The blue light time-series is shown in Figure
\ref{fig:bluetime}. Variation over each night is due to variable
slit losses and changing weather conditions.

\begin{figure}[htbp]
\vspace{9cm}
\center{\includegraphics{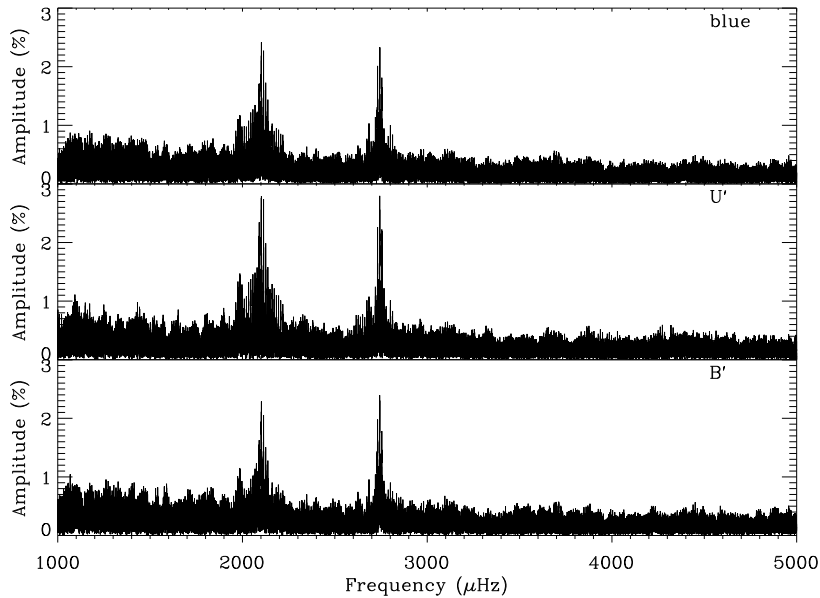}}
\caption{Amplitude spectra of the ``blue'' light curve (\textit{top
    panel}) and the $U^\prime$ and $B^\prime$ light curves.}
\label{fig:photspec}
\end{figure}

Figure \ref{fig:photspec} shows the amplitude spectra of each
time-series. There appears to be some amplitude variation between the
different passbands.

We fit the frequencies found by O'Toole et al. (2002a) to each
time-series and found that all modes we detected in the different
passbands are in phase. The ratios of $U^\prime$ to $B^\prime$
amplitudes are also the same for each mode, which is consistent with
them having the same $l$ value. We cannot, however, be certain of this
result, without detailed modelling.

\begin{figure}[htbp]
\vspace{8.5cm}
\center{\includegraphics{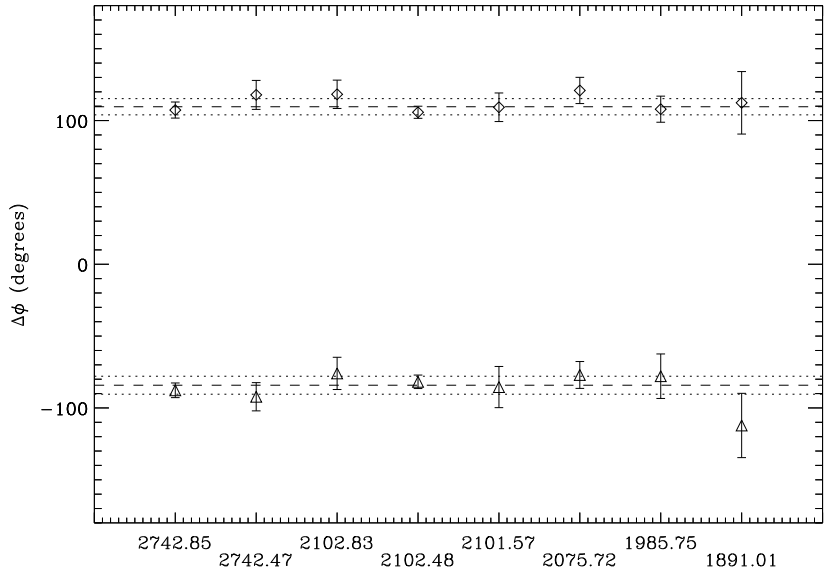}}
\caption{Phase differences between velocity and blue light photometry
for each mode found in velocity. The dashed line is the weighted
average phase difference and the dotted lines are one sigma errors.}
\label{fig:velhewph}
\end{figure}

\section{Comparison of equivalent width, spectrophotometry and
velocity}

The phase differences between velocity and intensity for each mode are
shown in the top panel of Figure \ref{fig:velhewph}, while the
difference between Balmer-line velocity and equivalent width are shown
in Figure \ref{fig:velhewph}. We calculated the average phase
difference between velocity and intensity variations
($\phi_\mathrm{vel}-\phi_\mathrm{phot}$) to be $109.6\pm5.6$
degrees. Positive velocity indicates redshift. For reference the Sun
has phase differences between $\sim$120 degrees and $\sim$130 degrees
dependent on frequency

The average phase difference between velocity and equivalent width
oscillations is $-84.2\pm6.3$ degrees. Therefore the equivalent width
variations are approximately 180 degrees out of phase with the
intensity variations, approximately what is expected - as the star
gets hotter, the equivalent width of the Balmer line decreases.
These phase differences are a measure of the degree of
non-adiabaticity of the oscillation in PG\,1605+072 and sdBs generally.

\acknowledgements
This work was supported by an Australian Postgraduate Award (SJOT),
the Australian Research Council, the Danish National Science Research
Council through its Center for Ground-based Observational Astronomy,
and the Danish National Research Foundation through its establishment
of the Theoretical Astrophysics Center.

\end{article}
\end{document}